\documentclass[a4paper]{article}

\usepackage[utf8]{inputenc}
\usepackage[english]{babel}
\usepackage{csquotes}
\usepackage[style=science, url=false, doi=false, backend=biber]{biblatex}
\usepackage{todonotes}
\usepackage{amsmath}
\usepackage{caption}
\usepackage{subcaption}
\usepackage{amsmath,amssymb,amsfonts,bm}
\usepackage[titletoc,title]{appendix}

\newcommand{\trt}{^\intercal}  
\newcommand{\Sint}{\bm{S}}  
\newcommand{\Lx}{L}  
\newcommand{\Nint}{N_\mathrm{x}}  
\newcommand{\Nsens}{N}  
\newcommand{\Nsrc}{M}  
\newcommand{\pinv}{^\dagger}  
\newcommand{\phivec}{\bm{\phi}}  
\newcommand{\noisevec}{\bm{n}}  
\newcommand{\leadfld}{\bm{\Gamma}}
\newcommand{\leadfldx}{\bm{\Gamma}_x}  
\newcommand{\avec}{\bm{a}}
\newcommand{\bsy}{\boldsymbol}

\author{Jussi Nurminen, Andrey Zhdanov, Wan Jin Yeo, Joonas Iivanainen, Julia Stephen, \\ Amir Borna, Jim McKay, Peter D.D. Schwindt, Samu Taulu}
\title{The effect of spatial sampling on the resolution of the magnetostatic inverse problem}

\bibliography{bibliography}

\begin{document}
\maketitle

\begin{abstract}

In magnetoencephalography, linear minimum norm inverse methods are commonly employed when a solution with minimal a priori assumptions is desirable. These methods typically produce spatially extended inverse solutions, even when the generating source is focal. Various reasons have been proposed for this effect, including intrisic properties of the minimum norm solution, effects of regularization, noise, and limitations of the sensor array. In this work, we express the lead field in terms of the magnetostatic multipole expansion and develop the minimum-norm inverse in the multipole domain. We demonstrate the close relationship between numerical regularization and explicit suppression of spatial frequencies of the magnetic field. We show that the spatial sampling capabilities of the sensor array and regularization together determine the resolution of the inverse solution. For the purposes of stabilizing the inverse estimate, we propose the multipole transformation of the lead field as an alternative or complementary means to purely numerical regularization.

\end{abstract}

\section{Introduction}

Non-invasive neuroimaging is the preferred method for investigation of anatomy and dynamics of the human brain, as invasive recordings are only feasible during certain clinical procedures. However, images obtained in non-invasive fashion are always limited to reconstruction based on measurements external to the head. Such reconstructions are prone to errors and imprecision due to e.g. measurement noise, instrument errors, and the assumptions made in the reconstruction algorithms.

Magnetoencephalography (MEG) \cite{hamalainen1993} allows the measurement of magnetic fields generated by neural currents with a high temporal resolution, but the spatial resolution of the reconstructed currents is limited by the mechanisms mentioned above \cite{lutkenhoner2003}. Interestingly, even in the absence of noise and instrumentation errors, the spatial resolution of the reconstructed images would still be limited by discrete spatial sampling of the magnetic field.

The sampling capabilities of multichannel MEG systems have been investigated from different points of view. Ahonen et al. \cite{Ahonen1993} considered the spatial aliasing of the recorded magnetic field in the presence of measurement noise. Kemppainen and Ilmoniemi \cite{Kemppainen1989} and Nenonen et al. \cite{Nenonen2004} applied total information based on Shannon's theory of communication to compare MEG sensor arrays. Vrba et al. \cite{vrba2004} and Tierney et al. \cite{Tierney2020} studied the effect of the number of channels on source localization.  Iivanainen et al. \cite{iivanainen2021} analyzed the effects of sampling on the spatial frequency content of the measured fields, and derived an information maximizing algorithm for optimizing sensor placement. 

For estimating the source distribution generating the MEG signal, methods based on the minimum norm estimate (MNE) \cite{hamalainen1994, Sarvas1987} are widely used. They are known to result in significant blurring of point-like source distributions, which may be quantified in terms of e.g. point-spread functions and spatial resolution or dispersion. Previous work has attributed the limited spatial resolution of the MNE inverse to various causes such as regularization, noise, the sampling provided by the sensor array, and the intrinsic properties of the inverse method \cite{deperaltamenendez1996, Hauk2004, Hauk2019, samuelsson2021, Molins2008}. However, the relative contributions of these factors remain unknown.

MNE-based methods depend on the properties of the MEG lead field matrix, which expresses the signal space topography of each elementary cortical source; the estimate is then a combination of these elementary sources \cite{hamalainen1994}.  In this work, we apply the magnetostatic multipole expansion \cite{Taulu2005} to transform the lead field from the sensor domain to the multipole domain, where field components are ordered hierarchically by their spatial frequency. In this way, the spatial properties of the lead field matrix can explicitly be controlled and related to the sampling capabilities of the sensor array. We contrast this approach with the traditionally employed numerical regularization of the lead field matrix, which corresponds to implicit spatial filtering.

\section{Multipole expansion of the magnetic field}

As a foundation, we describe the basic properties of the magnetostatic multipole expansion \cite{Taulu2005, Taulu2005b}. In neuromagnetic measurements, we can define a bounded region that contains the magnetic field sources of interest. Outside this region, assumed to be free of sources, the magnetic field can be expanded as

\begin{equation}
    \label{eq:field_decomp}
    \vec{B}(\vec{r}) = \sum_{l=1}^{\infty} \sum_{m=-l}^l \alpha_{lm}
    \frac{\vec{\nu}_{lm}}{r^{l+2}},
\end{equation}
where $\alpha_{lm}$ are the expansion coefficients, $\vec{\nu}_{lm}$ are modified vector spherical harmonics (VSH) \cite{Hill1954}, and $r$ is the distance from the chosen expansion origin. The integers $l$ and $m$ are the degree and order of the harmonic function, respectively. VSH terms with increasing $l$ represent progressively higher spatial frequencies.

Changing from continuous fields to a discrete sensor space representation, i.e. evaluating both sides of eq. \ref{eq:field_decomp} at a finite set of sensor locations, we obtain
\begin{equation}
    \label{eq:signalspace_decomp}
    \phivec = \sum_{l=1}^{\infty} \sum_{m=-l}^l \alpha_{lm} \bm{a}_{lm}.
\end{equation}
Here $\phivec$ is an $N$-dimensional signal space vector representing the field values measured by the $N$ sensors. $\bm{a}_{lm}$ are the $N$-dimensional multipole basis vectors, usually computed by integration of the continuous basis functions $\vec{\nu}_{lm} / r^{l+2}$ over the sensor geometries. If we also limit the sum to $l \leq \Lx$, we may express eq. \ref{eq:signalspace_decomp} in a matrix form as

\begin{equation}
    \phivec = \Sint \bm{x},
\end{equation}
where
\begin{equation}
\Sint  = \left[\avec_{1,-1}\ \avec_{1,0}\ \avec_{1,1}\ \ldots\ \avec_{\Lx,\Lx} \right] \\
\end{equation}

\begin{equation}
\bm{x}  = \left[\alpha_{1,-1}\ \alpha_{1,0}\ \alpha_{1,1}\ \ldots\ \alpha_{\Lx,\Lx} \right] \\ \trt.
\end{equation}

The estimated multipole components $\bm{\hat{x}}$ corresponding to a given magnetic field measurement $\phivec^\prime$ can then be obtained e.g. via the Moore--Penrose pseudoinverse as
\begin{equation}
\label{eq:multipole_est}
\bm{\hat{x}} = \Sint\pinv \phivec^\prime.
\end{equation}

In this way, the measurement is expressed as a linear superposition of field components oscillating at progressively higher spatial frequencies. Due to the $r^{-l-2}$ amplitude decay term in eq. \ref{eq:field_decomp}, components with higher $l$ value fall off faster with increasing distance from the source region. In practice, this means that components with increasingly high spatial frequencies become progressively weaker at the sensors. This is also the justification for truncating the representation of the signals to a finite value $l \leq L$, beyond which the components cannot be reliably measured due to insufficient signal-to-noise ratio (SNR).




Due to the typical geometry of the measurement, the VSH expansion is well suited for MEG. The expansion origin can be chosen so that all neural currents are located inside of a spherical volume whose radius is the distance from the expansion origin to the nearest sensor \cite{Taulu2005}.

\section{The inverse problem}

\subsection{The minimum norm pseudoinverse}

In the following, we denote the $\Nsens$-dimensional signal space vector comprising the signals from all sensors as $\phivec$, and the $\Nsrc$-dimensional vector comprising the amplitudes of all possible sources as $\bm{s}$. The sources have fixed orientation and location, and are typically obtained by discretization of the cerebral source volume. Due to the linear superposition principle, the total signal space vector resulting from the $\Nsrc$ sources can be written as

\begin{equation}
    \label{eq:leadfld}
    \phivec = \leadfld \bm{s}, 
\end{equation}
where $\leadfld$ is the $\Nsens \times \Nsrc$-dimensional lead field matrix, containing the signal space vectors corresponding to the individual sources. The lead field matrix depends on the source locations and orientations, as well as the choice of the forward model, such as a spherical or a boundary-element model.

The inversion of this forward equation seeks an estimate $\bm{\hat{s}}$ of the true source distribution $\bm{s}$. It is well known that the magnetic field outside the head, even if we are able to characterize it fully, gives us only limited information about the neuronal current distribution, and thus the inverse problem cannot be solved uniquely. Further, our measurement $\phivec$ is always limited by noise and the instrument sampling capabilities. 

Broadly, there are two categories of inverse methods: the first category  strongly restricts a priori assumptions on the source distribution (e.g. assuming a single focal source), while the second category does not. For the latter category of methods, the uniqueness of the inverse solution may be guaranteed in various ways, such as by minimizing the overall energy of the current distribution and the measured field. As pointed out in \cite{Hauk2004}, in the absence of a priori information and noise weighting, several such methods all reduce to the well-known Moore-Penrose pseudoinverse, i.e.

\begin{equation}
    \label{eq:sest}
    \bm{\hat{s}} = \leadfld\pinv\phivec = \leadfld\trt \left( \leadfld \leadfld\trt \right)^{-1} \phivec,
\end{equation}
where we assume that there are more sources than measurements ($M > N$). This solution has been termed the minimum-norm pseudoinverse (MNP).

\subsection{Regularization of the inverse}
\label{sec:regularization}

For realistic measurement and source geometries, the minimum-norm pseudoinverse of eq. \ref{eq:sest} turns out to be excessively sensitive to noise, necessitating regularization. The commonly employed Tikhonov regularized solution may be written as

\begin{equation}
    \label{eq:sest_tikh}
    \bm{\hat{s}} = \leadfld\trt \left( \leadfld \leadfld\trt + \lambda \bm{I} \right)^{-1} \phivec,
\end{equation}
where $\lambda$ is a regularization parameter to be determined. Another frequently used regularization method is the truncated singular value decomposition (TSVD). In fact, both methods can be expressed in terms of the singular value decomposition. Let $U \Sigma V\trt$ be the singular decomposition of $\leadfld$, where $U = \left[ \bsy{u}_1,  \ldots, \bsy{u}_\Nsens \right]$ is a $\Nsens \times \Nsens$ orthonormal matrix whose columns span the range of $\Lambda$, $\Sigma$ is a $\Nsens \times \Nsrc$ diagonal matrix of singular values $\sigma_i$ and $V = \left[ \bsy{v}_1,  \ldots, \bsy{v}_\Nsrc \right]$ is a $\Nsrc \times \Nsrc$ orthonormal matrix. The pseudoinverse of eq. \ref{eq:sest} may then be written as

\begin{equation}
\label{eq:sest_pinv_svd}
\bm{\hat{s}} =\sum_{i=1}^N \frac{\bsy{u}_i\trt \phivec}{\sigma_i} \bsy{v}_i.
\end{equation}

%

In the TSVD, the terms corresponding to the smallest singular values are eliminated by truncating the sum, i.e.

\begin{equation}
\label{eq:sest_tsvd}
\bm{\hat{s}}_{\mathrm{TSVD}}=\sum_{i=1}^K \frac{\bsy{u}_i\trt \phivec}{\sigma_i} \bsy{v}_i,
\end{equation}
where the regularization parameter $K$ represents the truncation point. On the other hand, the Tikhonov regularized solution can be shown to equal a weighted version of the sum:

\begin{equation}
\label{eq:tsvd}
\bm{\hat{s}}_{\mathrm{Tikhonov}}=\sum_{i=1}^N f_i \frac{\bsy{u}_i\trt \phivec}{\sigma_i} \bsy{v}_i,
\end{equation}
where the weighting factors $f_i = \frac{\sigma_i^2}{\sigma_i^2 + \lambda^2}$ depend on the regularization parameter $\lambda$: terms corresponding to singular values significantly smaller than $\lambda$ will be suppressed. The effect of both regularization methods is to suppress the contribution of terms corresponding to small singular values, which suffer from the largest inaccuracy.

A disadvantage of numerical regularization methods is the somewhat arbitrary choice of the regularization parameter. The regularization parameter $\lambda$ can be related to the assumed measurement SNR, but nevertheless is still a free parameter. The Tikhonov regularized solution (eq. \ref{eq:sest_tikh}) can be viewed as a special case of the generalized MNE where the source and noise covariance matrices are proportional to diagonal matrices i.e. the source amplitudes and noise at the sensors are independent Gaussian random variables with variances $a^2$ and $b^2$. In this case the regularization parameter $\lambda$ is related to their ratio: $\lambda = b^2/a^2$, the inverse of the assumed SNR. MNP (eq. \ref{eq:sest}), on the other hand, can be viewed as the limiting case of the Tikhonov regularized solution in the limit of infinite SNR ($a^2/b^2 \to \infty)$.

\subsection{The lead field in the multipole domain}

For traditional MEG sensor array geometries, the degrees of freedom obtainable from the measurement are considerably fewer than the number of sensors. Thus, the sensor-based lead field has significant redundancy. As an alternative, the lead field may also be expressed in the multipole domain. This may be accomplished either by directly computing the multipole component for each element of the source space, or alternatively by first computing the conventional sensor-based lead field and then transforming it according to

\begin{equation}
    \label{eq:leadfld_multipole}
    \leadfldx = \Sint\pinv \leadfld.
\end{equation}
The columns of $\leadfld$, i.e. the forward fields, are thus expressed in terms of multipole components, rather than sensor amplitudes. The dimension of $\leadfldx$ is $\Nint \times \Nsrc$, where $\Nint$ is the number of multipole components which can be chosen to express the essential degrees of freedom contained in the measurement, typically $\Nint \ll \Nsens$. Thus, $\leadfldx$ is a more economical description of the source forward fields, while retaining all information obtainable by the sensor array. This is reflected by the typically significantly lower condition number of $\leadfldx$.

Analogously to eq. \ref{eq:sest}, a source estimate can be obtained via the multipole-domain lead field as
\begin{equation}
    \label{eq:sest_multipole}
    \bm{\hat{s}} = \leadfldx\pinv\bm{x},
\end{equation}
where $\bm{x}$ are the multipole coefficients corresponding to a measurement, typically obtained by eq. \ref{eq:multipole_est}. This form could be directly used as a source estimate. However, we note that inserting eqs. \ref{eq:multipole_est} and \ref{eq:leadfld_multipole} will lead to

\begin{equation}
    \bm{\hat{s}} = (\Sint\pinv\leadfld)\pinv\Sint\pinv\phivec = \leadfld\pinv\Sint\Sint\pinv\phivec.
\end{equation}

Since $\Sint\Sint\pinv$ is an orthogonal projection operator, the multipole domain source estimate is mathematically equivalent to first projecting $\phi$ onto the range of $\Sint$ and subsequently applying the conventional pseudoinverse of eq. \ref{eq:sest}. However, from the numerical point of view, $\leadfldx$ typically has better inversion properties and thus it is advantageous to use it instead of $\leadfld$.

Finally, we note that it is common to restrict the sensor-based inverse solution of eq. \ref{eq:sest} to use a subset of sensors, e.g. in the case where some sensors are noisy or malfunctioning, or when a solution corresponding to a region of interest is desired. Similarly, we have the freedom to choose which multipole components to use in the inverse. Thus we can use a subset of $\bm{x}'$ of the estimated multipole coefficients and limit the lead field to these components, resulting in 

\begin{equation}
    \label{eq:sest_multipole_filtered}
    \bm{\hat{s}} = \leadfldx^{\prime \ \dagger} \bm{x}'.
\end{equation}

For example, we might choose to only include spatial frequencies up to a certain threshold, or weight the components according to SNR.

\subsection{The resolution matrix and its properties}

For a general linear inverse method, the source estimate may be written

\begin{equation}
    \label{eq:lin_sest1}
    \bm{\hat{s}} = \bm{M} \phivec,
\end{equation}
where $\bm{M}$ is the linear estimator. Inserting eq. \ref{eq:leadfld}, we obtain

\begin{equation}
    \label{eq:lin_sest2}
    \bm{\hat{s}} = \bm{M} \leadfld \bm{s} \equiv \bm{\Omega} \bm{s}.
\end{equation}
where $\bm{\Omega} = \bm{M} \leadfld$ may be termed the resolution matrix, first applied in the context of biomagnetic inverse problems in \cite{deperaltamenendez1996}.  This equation expresses the source estimate as a weighted combination of columns of $\bm{\Omega}$. Thus, the columns can be interpreted as unit estimates, or equivalently as point-spread functions (PSFs) corresponding to each source. On the other hand, the estimate $\bm{\hat{s}}(k)$ for the $k$th source is

\begin{equation}
    \label{eq:lin_sest}
    \bm{\hat{s}}(k) = \sum_j \bm{\Omega}_{kj} s_j,
\end{equation}
where $\bm{\Omega}_{ij}$ are the elements of the resolution matrix. From this form, it is seen that the $k$th row of the matrix describes the (undesirable) contribution of other sources to $\bm{\hat{s}}(k)$. In previous literature, the rows have been correspondingly termed resolution kernels, or sometimes cross-talk functions (CTFs) \cite{deperaltamenendez1996, Hauk2004, Hauk2019, samuelsson2021}.

For the simple minimum norm pseudoinverse,

\begin{equation}
    \label{eq:reskernel}
    \bm{\Omega} = \leadfld \pinv \leadfld,
\end{equation}
i.e., the resolution matrix is symmetrical, and thus the PSFs and CTFs for a given source are identical. This also holds for the Tikhonov regularized version of eq. \ref{eq:sest_tikh}, in which case we get

\begin{equation}
    \label{eq:reskernel_tikhonov}
    \bm{\Omega} = \leadfld\trt \left( \leadfld \leadfld\trt + \lambda \bm{I} \right)^{-1} \leadfld.
\end{equation}

The spatial extent of resolution matrix PSFs is of interest, since it determines the spatial blurring of the total source estimate. It has previously been quantified in terms of PSF spatial dispersion (SD), with slightly differing definitions \cite{samuelsson2021, hauk2011, Molins2008, hedrich2017}. Here we define it for the $k$th source as
\begin{equation}
    \label{eq:spatial_dispersion}
    \mathrm{SD}(k) = \sqrt{\frac{\sum_{i=1}^M d_{ik}^2 \bm{\Omega}_{ik}^2} {\sum_{i=1}^M \bm{\Omega}_{ik}^2}},
\end{equation}
where $d_{ik}$ is the Euclidian distance between source nodes $i$ and $k$.

Finally, as noted in the previous section, we can also apply the inverse in the multipole domain. Inserting eq. \ref{eq:leadfld_multipole}, we obtain a direct expression for the multipole-based resolution matrix as

\begin{equation}
    \label{eq:reskernel_multipole}
    \bm{\Omega}_{x} = \leadfldx\pinv\leadfldx = \leadfld\pinv(\Sint\pinv)\pinv\Sint\pinv\leadfld=\leadfld\pinv\Sint\Sint\pinv\leadfld,
\end{equation}

where we can also apply Tikhonov regularized inversion similarly to eq. \ref{eq:reskernel_tikhonov}, if necessary.

\section{Results}

\subsection{Effect of spatial frequencies on the minimum norm solution}

We computed the MNE inverse solutions for forward fields of single sources in a single-layer boundary element model using the MNE-Python software package \cite{Gramfort2013, Gramfort2014}. The sources (N=7498) were placed approximately uniformly on the MRI-derived cortical surface of the MNE-Python \emph{sample} subject, using subsampling of the cortical surface. The source orientations were constrained according to the orientation of the local cortical surface normal. To represent the measurement of the magnetic fields, the radial components of the forward fields were computed at 1000 locations on a full spherical surface (R=120 mm) around the head. Note that compared to typical MEG sensor arrays, this array represents a relatively high fidelity of spatial sampling. We used it here to study the limits of attainable spatial resolution, without being significantly limited by the sampling of the magnetic field.

From the computed lead fields, we determined the sensor-based and multipole-based resolution matrices according to eqs. \ref{eq:reskernel} and \ref{eq:reskernel_multipole}. The multipole-based matrices were computed for maximum expansion degrees $L = 1-16$. The PSFs are the columns of the resolution matrices. The spatial dispersion values were computed from the PSFs according to eq. \ref{eq:spatial_dispersion}.

\begin{figure}[htb!]
    \centering
    \includegraphics[width=\textwidth]{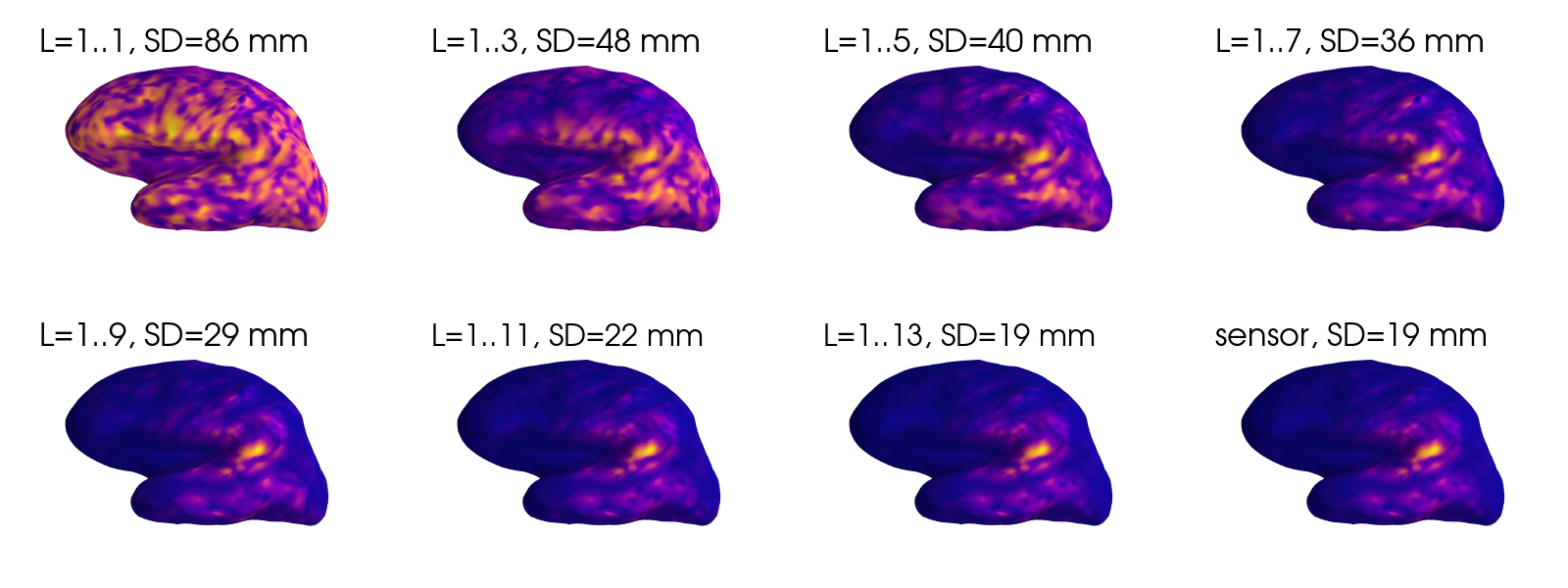}
    \caption{Point spread functions of the multipole and sensor based inverses for a representative superficial source. Note that due to the differing peak magnitudes of the PSFs, the plots are scaled individually according to their respective maxima. $L$ indicates the spherical harmonics degrees used for computation of the multipole-based resolution matrices. SD indicates the spatial dispersion of the point spread function.}
    \label{fig:psf_vs_L}
\end{figure}

Figure \ref{fig:psf_vs_L} shows the point spread function for the multipole-based inverse with different $L$ values, as well as for the sensor-based inverse. As higher spatial frequencies are included in the inverse, the PSF becomes more focal, converging towards the sensor-based inverse which includes all spatial frequencies. To facilitate comparison, all the PSFs were computed using Tikhonov regularization with $\lambda=10^{-11}$.

\begin{figure}[htb!]
    \centering
    \includegraphics[width=\textwidth]{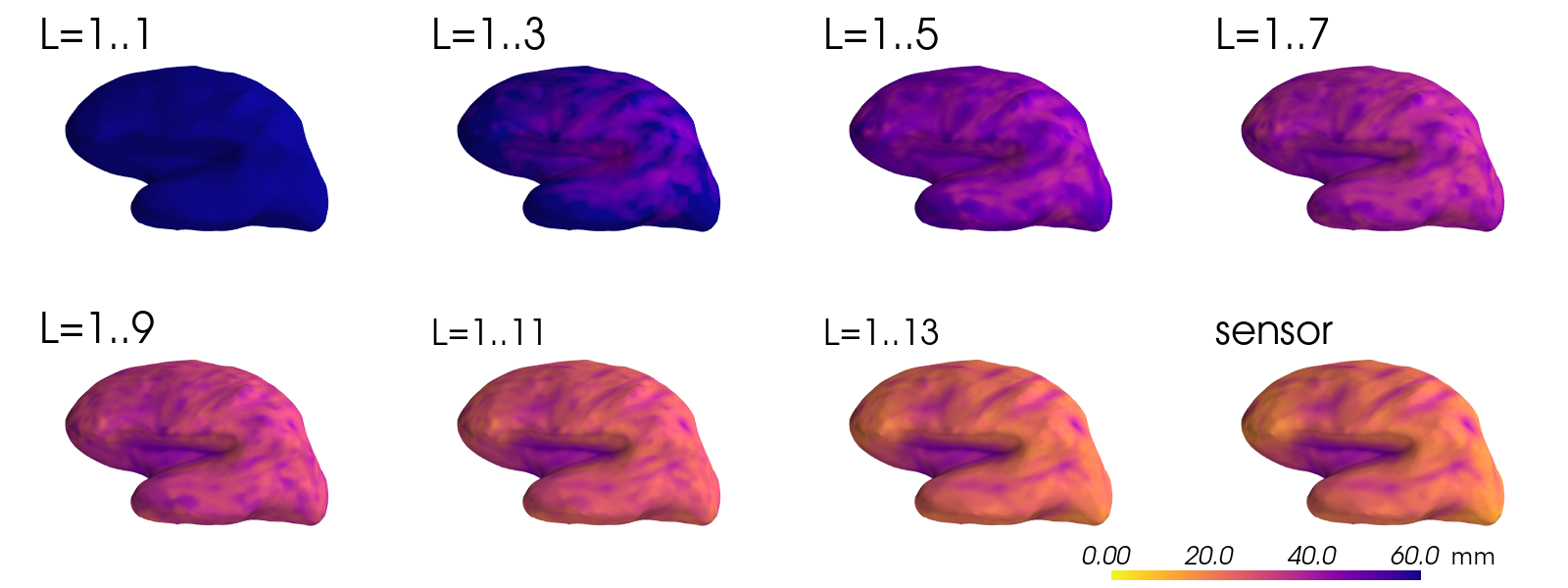}
    \caption{Spatial dispersion of the resolution matrix PSFs as a function of source location. $L$ indicates the spherical harmonics degrees used for computation of the multipole-based resolution matrices. The plots are all equally scaled.}
    \label{fig:dispersion_vs_L}
\end{figure}

Figure \ref{fig:dispersion_vs_L} shows the spatial dispersion of the point spread function for different cortical locations, as a function of the multipole expansion order. In accordance with figure \ref{fig:psf_vs_L}, spatial dispersion is reduced with increasing $L$ cutoff and converges towards the sensor space result as higher spatial frequencies are included in the inverse. As in the previous step, Tikhonov regularization with $\lambda=10^{-11}$ was used.

\subsection{Effect of regularization on the spatial frequency spectrum}

\begin{figure}[htb!]
    \centering
    \includegraphics[width=\textwidth]{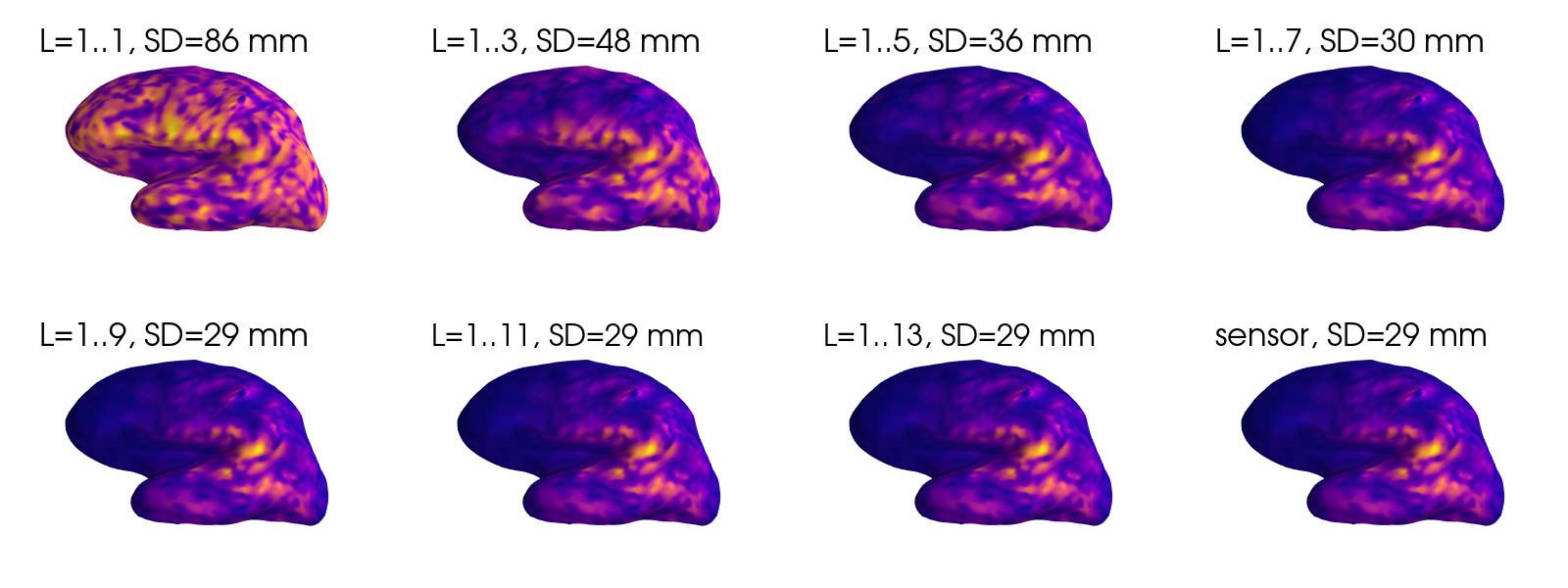}
    \caption{Point spread functions of the multipole and sensor based inverses for a representative superficial source for $\lambda=10^{-8}$. The plots are scaled individually according to their respective maxima.}
    \label{fig:psf_vs_L_strong_regu}
\end{figure}

To illustrate the effect of the regularization parameter, we recomputed the PSFs of figure \ref{fig:psf_vs_L} with $\lambda=10^{-8}$, as shown in figure \ref{fig:psf_vs_L_strong_regu}. It is evident that including spatial frequencies beyond about $L=7$ no longer improves the focality of the solution, since the stronger regularization effectively eliminates these frequencies.

\begin{figure}
    \centering
    \includegraphics[width=\textwidth]{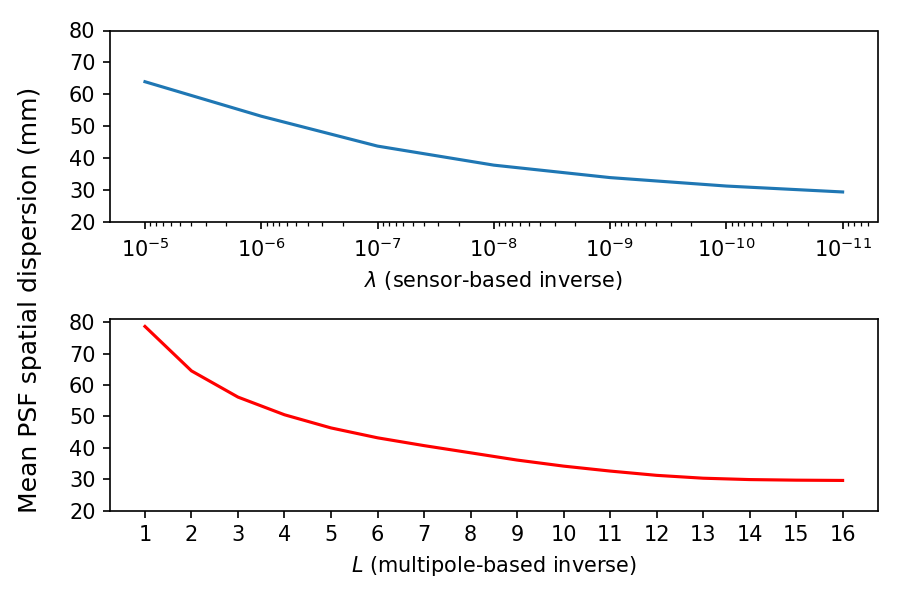}
    \caption{Mean spatial dispersion of resolution matrix PSFs over the whole cortical source space, as a function of $\lambda$ (for sensor-based inverse) and L (for multipole-based inverse). For multipole based inverse, no numerical regularization was applied.}
    \label{fig:SD_vs_L_and_lambda}
\end{figure}

The effect of numerical regularization vs.\ $L$-filtering of the lead field is further illustrated in figure \ref{fig:SD_vs_L_and_lambda}. Reducing the amount of numerical regularization has an effect very similar to increasing $L$.

\subsection{The singular value decomposition of the lead field and the magnetostatic multipole components}

\begin{figure}
    \centering
    \includegraphics[width=\textwidth]{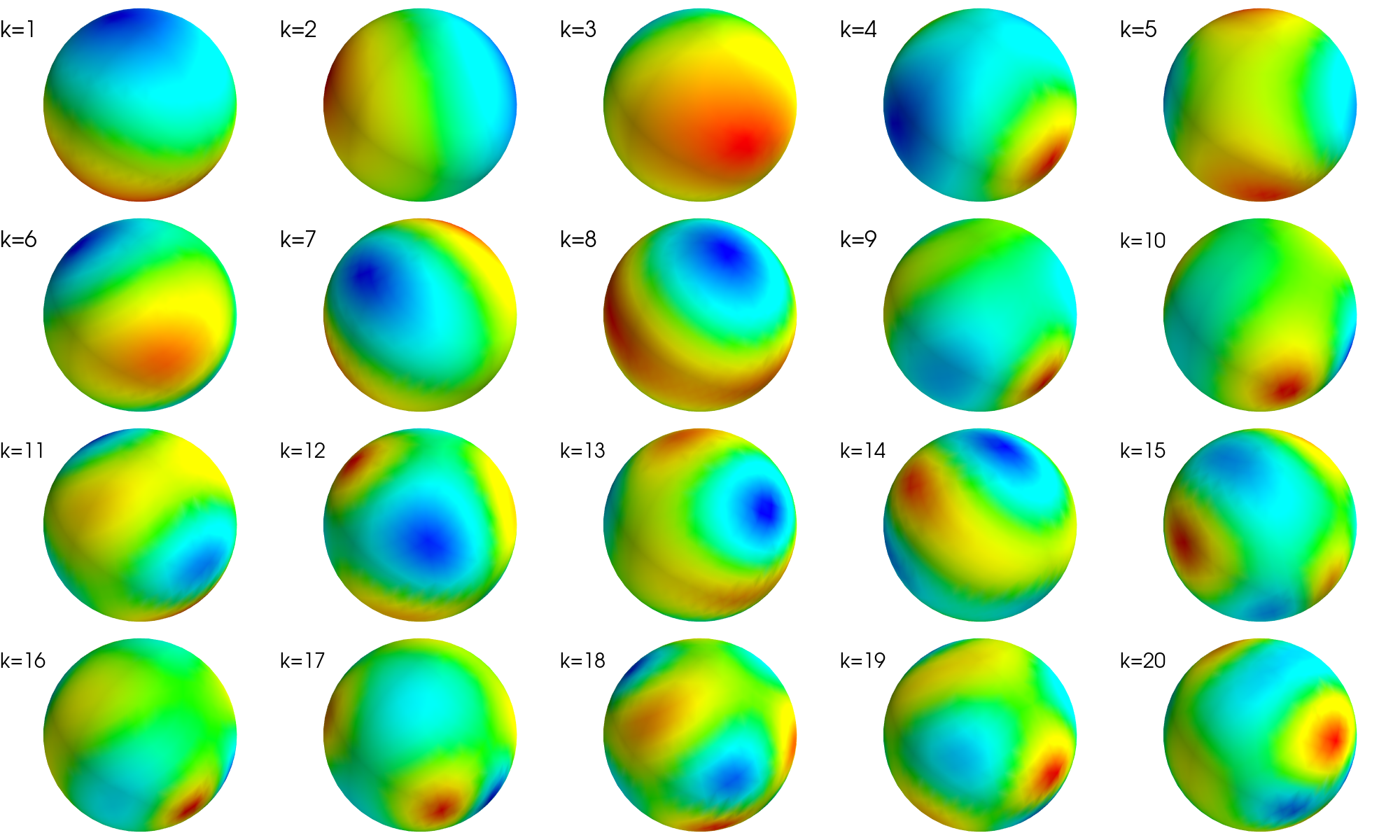}
    \caption{Left-hand singular vectors $\bsy{u}_i$ of the lead field matrix, corresponding to the 20 largest singular values. The plots are individually scaled.}
    \label{fig:svd_basis_trimesh}
\end{figure}

According to section \ref{sec:regularization}, the lead field is spanned by its left-side singular vectors $\bsy{u}_i$. They may be viewed as elementary forward fields, from which any measurable field can be built. On the other hand, we have previously shown that any field can also be expressed in terms of the VSH basis vectors $\bsy{a}_{L,m}$. In fact, in turns out that $\bsy{u}_i$ and $\bsy{a}_{L,m}$ are highly similar. Figure \ref{fig:svd_basis_trimesh} illustrates the first 20 $\bsy{a}_{L,m}$ and $\bsy{u}_i$ vectors for our measurement geometry. It is seen that decreasing singular values and increasing $L$ values correspond to increasingly high spatial frequencies. These high frequencies decay the fastest and produce the weakest signals at the sensors.

Accordingly, there is a close relationship between the traditional Tikhonov or TSVD regularized inverse solution, and the solution based on the the multipole-domain lead field. In the former, spatial frequencies are implicitly limited by the regularization parameter, which determines the SVD components included in the inverse. In the latter case, the spatial frequencies are explicitly determined by the VSH degree cutoff $L$.

\begin{figure}
    \centering
    \includegraphics[width=\textwidth]{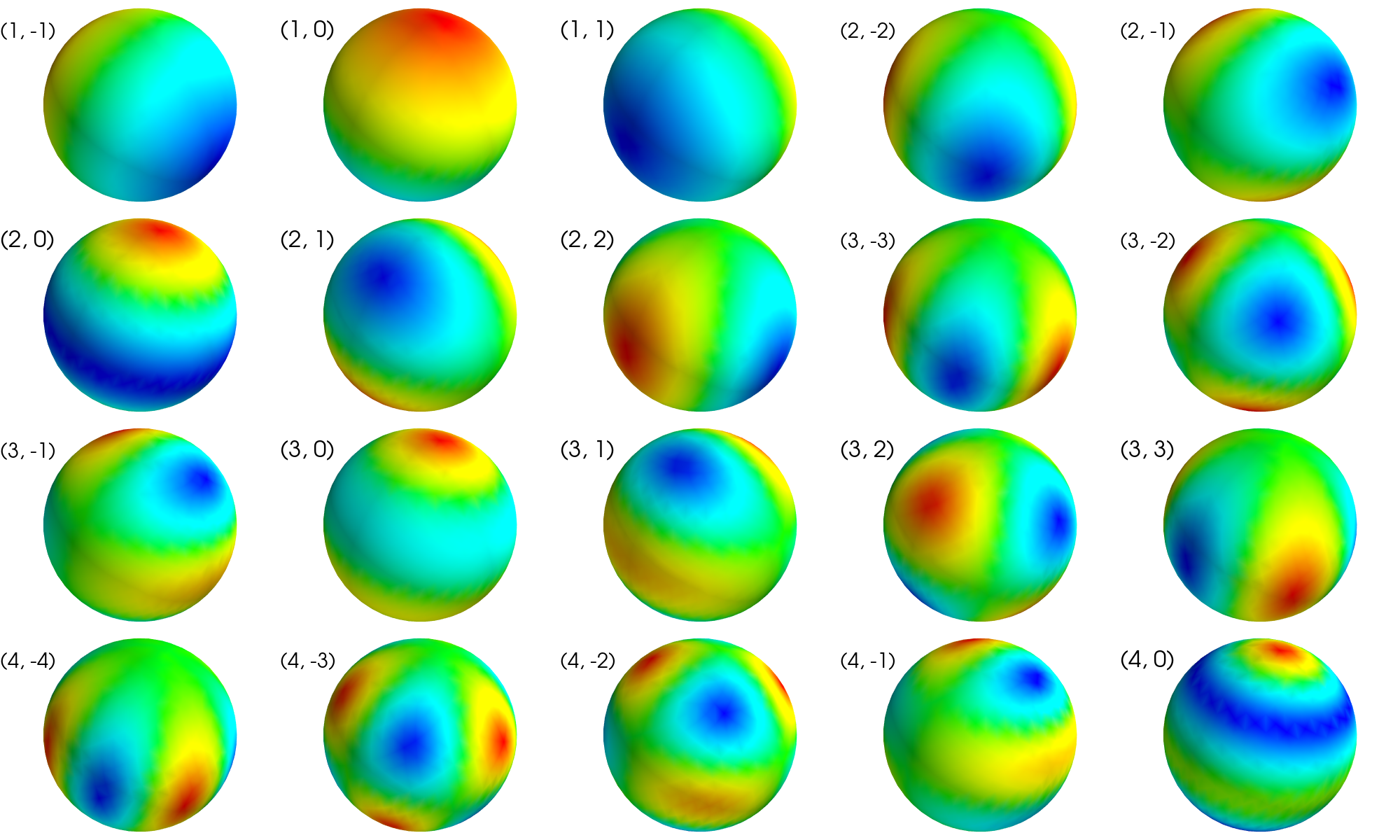}
    \caption{First 20 vector spherical harmonic (VSH) basis functions $\bsy{a}_{L,m}$ corresponding to the lowest spatial frequencies. The captions indicate the (L, m) values for each basis vector. The plots are individually scaled.}
    \label{fig:vsh_basis_trimesh}
\end{figure}

\subsection{The multipole transformation as physics-based regularization and the effect of SNR}

From the above discussion, it follows that we can use elimination of high spatial frequencies ("$L$-filtering") as a regularization method. If the lead field is limited to relatively low spatial frequencies, further numerical regularization may not be necessary. Thus, the multipole-based transformation of the lead field offers an alternative, physics-based method of regularizing the lead field matrix.

To demonstrate this, we selected the same source whose PSF is illustrated in Fig. \ref{fig:psf_vs_L} and added uniform Gaussian noise to its forward field. Here we define the SNR as the ratio of signal vector norms
\begin{equation}
    \mathrm{SNR} = \|\phivec\| / \|\noisevec\|,
\end{equation}
where $\phivec$ is the signal space vector corresponding to the source, and $\noisevec$ is a realization of Gaussian noise.

Next, we performed the inverse in the multipole domain for various $L$ values. The noisy multipole-domain signal is computed as

\begin{equation}
    \bm{x}_n = \Sint\pinv (\phivec + \noisevec)
\end{equation}

and the lead field as

\begin{equation}
    \leadfldx = \Sint\pinv \leadfld.
\end{equation}

The included spatial frequencies are determined by the choice of $L$, which in turn determines $\Sint$. The source estimate is then obtained by 

\begin{equation}
    \bm{\hat{s}} = \leadfldx\pinv\bm{x}_n.
\end{equation}

Note that here the direct pseudoinverse of $\leadfldx$ is used (i.e. no numerical regularization) to evaluate the efficiency of $L$-filtering. 

\begin{figure}
    \centering
    \includegraphics[width=\textwidth]{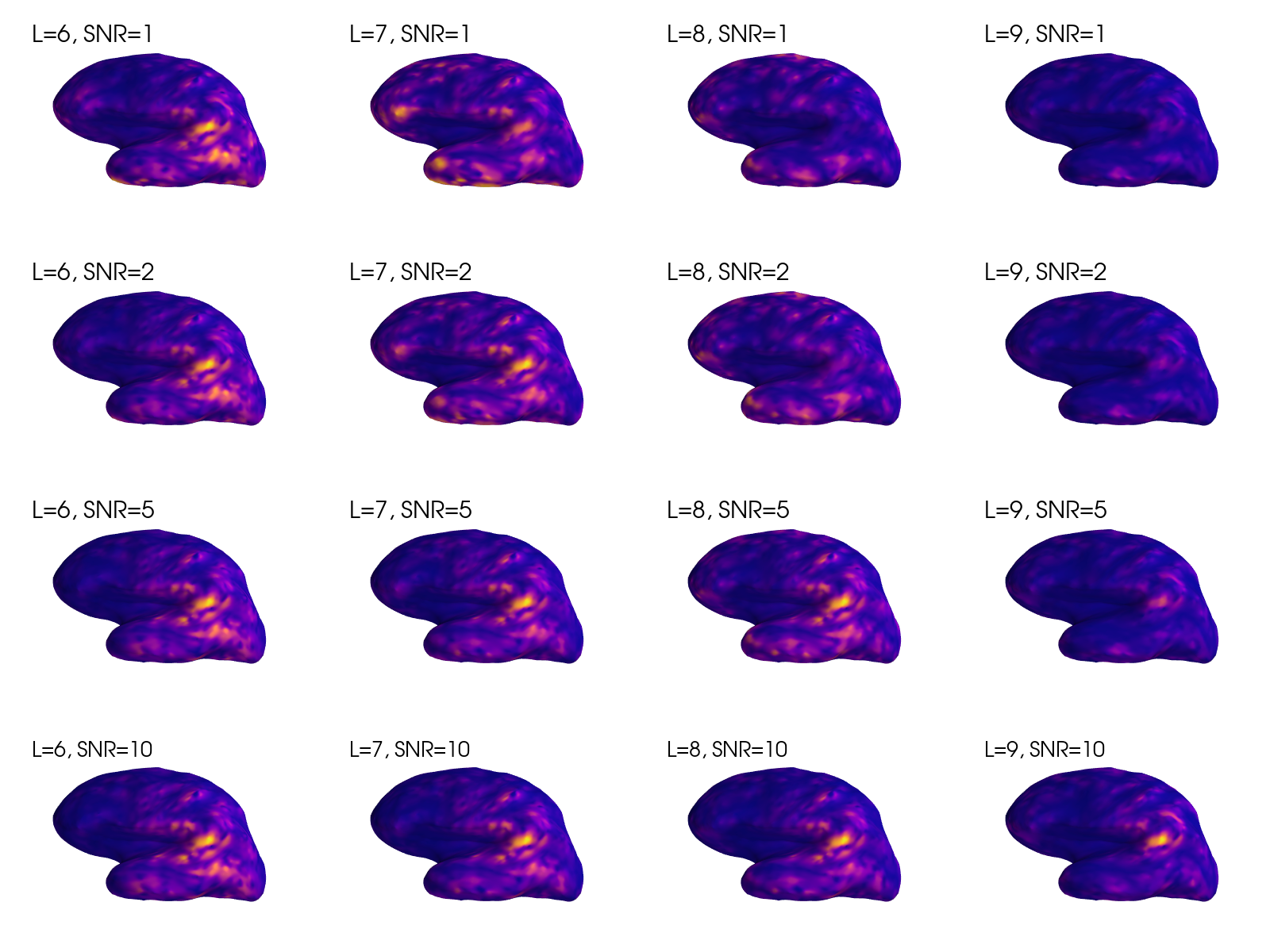}
    \caption{Inverse solutions without numerical regularization for a single source as a function of $L$ and SNR. Plots are individually scaled according to their respective maxima. Some solutions appear very small in amplitude despite scaling, which is caused by an invisible spurious peak outside the displayed cortical region. To remove the effect of random noise topography, the same noise realization was used for the different SNR values, scaled according to SNR.}
    \label{fig:inverse_vs_SNR_and_L}
\end{figure}

The results are shown in fig. \ref{fig:inverse_vs_SNR_and_L}. We note that $L=6$ yields a reasonable solution for all SNR values, though the focality is limited. $L=8$ requires a SNR of 5 for a reasonable solution. For $L=9$, SNR=10 is needed for an accurate solution, but the resulting solution is quite focal. In summary, inverse solutions limited to low $L$ values are better conditioned, producing reasonable results even in the case of low SNR. However, due to the limited spatial frequencies, they are unable to reproduce focal sources.

\section{Discussion}

The effect of conventional numerical regularization on MNE-type inverse solutions is to impose a limit on the spatial frequencies contained in the solution. The choice of regularization parameter sets an ultimate bound on the attainable focality of the inverse solution, by suppressing high spatial frequencies that are needed to fully characterize focal source distributions. On the other hand, noise imposes a practical bound on the regularization parameter, since insufficient regularization in case of noisy data will lead to spurious solutions.

The magnetostatic multipole expansion provides an economical and spatially organized description of the measured fields, which has well-established applications in interference suppression and movement compensation \cite{Taulu2005, taulu2009}. Its potential uses in source modelling have remained largely unexplored so far; however see \cite{vrba2010}, where the multipole expansion was utilized in the context of beamformers. Here we demonstrate the utility of multipole-based signal representation for inverse solutions that rely on lead field inversion. It is advantageous to limit the modelled forward fields, in this case the lead field, to the spatial frequencies that can be reliably detected by the instrument. To this end, we have demonstrated that choosing a suitable spatial frequency cutoff and performing the inverse in the multipole domain may eliminate the need for numerical regularization. Thus, the multipole transformation may be interpreted as a physics-based regularization method that allows the exclusion of high spatial frequencies in an exact way.

Any sensor array has a practical upper limit $L$ of spatial frequencies that it is capable of measuring. The limit is determined not only by the sampling capabilities of the array and its calibration accuracy but also by its noise level, since high spatial frequencies are weakest at the sensors and beyond a certain limit will be buried in noise. For example, in the case of the commercially available 306-channel MEGIN (formerly Elekta) TRIUX instrument, it has been shown that $L=8$ provides sufficient characterization even for the most superficial source configurations. Thus, the MEGIN MaxFilter software by default limits the signal spatial frequencies to $L=8$. Our results indicate that the additional resolution provided by high spatial frequencies such as $L>10$ would require extremely high SNR values, which are not obtainable in standard MEG studies with human subjects. However, emerging sensor arrays based on optically pumped magnetometers are likely to benefit from the inclusion of higher spatial frequencies in source modelling, due to the proximity of the sensors to the head and the correspondingly higher SNR.

In future studies, it would also be desirable to explore inverse solutions optimally weighted by spatial frequency, or more broadly, by the degree and order of the multipole basis components. For example, the sensitivity of a sensor array to different multipole components varies according to array geometry; thus, giving more weight to components with the highest signal-to-noise would be expected to yield improved results in source reconstruction. 

\section{Acknowledgements}

This work was supported by grants R21-EB033577-01 and U01-EB028656-04 from the National Institutes of Health.

S. Taulu's work is also funded in part by the Bezos Family Foundation and the R. B. and Ruth H. Dunn Charitable Foundation. 

Sandia National Laboratories is a multimission laboratory managed and operated
by National Technology \& Engineering Solutions of Sandia, LLC, a wholly owned
subsidiary of Honeywell International Inc., for the U.S. Department of Energy
National Nuclear Security Administration under contract DENA0003525. This paper
describes objective technical results and analysis. Any subjective views or
opinions that might be expressed in the paper do not necessarily represent the
views of the U.S. Department of Energy and the United States Government. The content is solely the responsibility of
the authors. 

\printbibliography

\end{document}